\documentclass[
 amsmath,amssymb,
preprint,
]{revtex4-1}

\usepackage{graphicx}
\usepackage{dcolumn}
\usepackage{bm}
\usepackage{textcomp}
\usepackage{gensymb}
\usepackage{color,soul}

\usepackage[utf8]{inputenc}

\usepackage[version=4]{mhchem}
\usepackage{hyperref}
\begin{document}

\preprint{APLMat/OSULLIMBRO}

\title{Epitaxial growth, optical and electrical conductivity of the metallic pyrochlore \ce{Bi2Ru2O7} on Y-stabilized \ce{ZrO2} substrate.}

\author{Marita O'Sullivan}
\email{mosulliv@liverpool.ac.uk}
\affiliation{Department of Physics, University of Liverpool, Oxford Street, Liverpool, L69 7ZE, United Kingdom}%
\author{Matthew S. Dyer}
\affiliation{
Department of Chemistry, University of Liverpool, Crown Street, Liverpool, L69 7ZD, United Kingdom
}
\author{Michael W. Gaultois}
\affiliation{
Department of Chemistry, University of Liverpool, Crown Street, Liverpool, L69 7ZD, United Kingdom
}
\affiliation{
Leverhulme Research Centre for Functional Materials Design, The Materials Innovation Factory, University of Liverpool, 51 Oxford Street, Liverpool, L7 3NY, United Kingdom
}

\author{John B. Claridge}
\affiliation{
Department of Chemistry, University of Liverpool, Crown Street, Liverpool, L69 7ZD, United Kingdom
}
\author{Matthew J. Rosseinsky}
\email{m.j.rosseinsky@liverpool.ac.uk}
\affiliation{
Department of Chemistry, University of Liverpool, Crown Street, Liverpool, L69 7ZD, United Kingdom
}
\author{Jonathan Alaria}%
\affiliation{Department of Physics, University of Liverpool, Oxford Street, Liverpool, L69 7ZE, United Kingdom}%

\emph{The following article has been submitted to APL Materials. After it is published, it will be found at \href{https://publishing.aip.org/resources/librarians/products/journals/}{Link}.}

\date{\today}

\begin{abstract}
Epitaxial heterostructures composed of complex correlated metal oxides grown along specific crystallographic orientations offer a route to investigating emergent phenomena such as topological states and spin liquids through geometrical lattice engineering. A$_2$Ru$_2$O$_7$ pyrochlore ruthenates in particular exhibit a metal-insulator transition with varying A cation whose mechanism is not fully understood. We report on the epitaxial growth, structural and electrical properties of metallic pyrochlore bismuth ruthenate heterostructures grown along both the [001] and [111] directions. Ordered pyrochlore thin films were obtained with highly oriented texture along the [001] and [111] crystallographic directions. Density functional theory calculations of the electronic band structure and density of states indicated that Bi$_2$Ru$_2$O$_7$ is semimetallic and that hybridization of the Ru 4$d$ and Bi 6$p$ orbitals via the anion network at the Fermi energy was responsible for the metallicity. Electrical conductivity measurements confirmed that the compound is weakly metallic in agreement with the reported conductivity for the stoichiometric bulk compound. The carrier concentration and mobility of the electrons compared favorably with previous reports on bulk material and indicate strong electron-electron interactions. The measured and computed optical conductivities were found to share coincident spectral features and confirm the electronic correlation. Comparison of the electrical and optical properties of the two distinct orientations indicates differences that cannot be attributed to differences in crystalline quality or dislocations and may indicate anisotropy in the electronic structure of Bi$_2$Ru$_2$O$_7$. This study will enable access to the kagome lattice arising naturally in the {111} planes of the pyrochlore B cation sublattice which may be used to uncover emergent topological properties.

\end{abstract}

\maketitle

\section{Introduction}
Oxides with the statistical formula \ce{A2B2O7} crystallizing with the pyrochlore structure have shown wide-ranging properties suitable for applications such as oxide ion conductivity\cite{Li20131296}, metallic conductivity as thick film electrodes\cite{Zhang2021}, ferroelectricity\cite{Dong2009}, and thermoelectricity\cite{Yasukawa2003213,Akhbarifar2020}. They also offer a rich playground to explore more fundamental physical properties with potential topological materials\cite{Pesin2010376,Yang2014,Chu2019,PhysRevLett.103.206805,FieteSciRep2015} and canonical spin ice behavior\cite{Bovo2014}.

The \ce{A2Ru2O7} pyrochlore ruthenates in particular exhibit electrical conductivity that verges between localized and itinerant behavior and has led to numerous studies into the evolution of their electronic structures with varying A cation to understand the mechanism of the Mott metal-insulator transition. \ce{Bi2Ru2O$_{7-y}$} and \ce{Pb2Ru2O$_{6.5}$}, for example, are weakly metallic Pauli paramagnets\cite{Yamamoto1994372,Kobayashi199515}, while \ce{Y2Ru2O7} and \ce{Ln2Ru2O7} (Ln = Pr--Lu) exhibit semiconducting behavior with spontaneous antiferromagnetic order\cite{Bouchard1971669,Gokagac1994235,Gaultois2013}. Electrical conduction in the metallic ruthenates is known to take place via the \ce{Ru-O} band states in the \ce{Ru2O6} network\cite{LEE1997405}. In photoemission spectroscopy and high-resolution electron energy-loss spectroscopy studies, Cox $et$ $al.$ identified a decrease in the density of states at the Fermi energy, $E_{\textnormal{F}}$, from \ce{Pb2Ru2O$_{6.5}$} to \ce{Bi2Ru2O7} to \ce{Y2Ru2O7} with electron-electron correlation yielding a Mott insulator for \ce{Y$_2$Ru$_2$O$_7$}\cite{Cox19836221}. They posited that mixing of the A 6$s$ and Ru 4$d$ bands via the A--O--Ru interaction had the effect of broadening the Ru $t_{2g}$ block bands arising from the octahedral crystal field \cite{Cox19836221}, however changes to the bandwidth or band splittings can be observed as a function of the A cation. It was noted that increasing the A site radius has the effect of opening the \ce{Ru-O-Ru} bond angle thereby increasing the Ru 4$d$/O 2$p$ orbital overlap and broadening the bandwidth to delocalize the carriers\cite{LEE1997405}. It was therefore concluded that metallic ruthenate pyrochlores are characterized by relatively short \ce{Ru-O} bonds ($\sim$1.95 \AA), less distorted \ce{RuO6} octahedra, and more open \ce{Ru-O-Ru} angles ($>$ 133\degree) compared with longer bonds ($\sim$1.98 \AA), greater octahedral distortions, and smaller angles ($<$ 133\degree) in the semiconducting analogues\cite{Facer19931897,Kanno1993106,Kennedy1996261,Okamoto2004,LEE1997405}. 

Self-consistent first principles calculations of the electronic band structure of metallic pyrochlore ruthenates using the pseudofunction method, however, were used to argue that the \ce{A} 6$s$ bands are too deep to mix with \ce{Ru} 4$d$ bands\cite{Hsu1988792}. Local-density approximation density functional theory calculations suggested that the unoccupied \ce{A} 6$p$ states positioned slightly above the Fermi energy, $E_{\textnormal{F}}$, could hybridize with the \ce{Ru} 4$d$ $t_{2g}$ states via the \ce{O} 2$p$ and contribute band filling charge resulting in metallic conductivity\cite{Ishii2000526,PhysRevB.72.035124}. Indeed the position of the \ce{A} cation orbitals relative to the \ce{Ru} $d$\ce{-}\ce{O} $p$ antibonding states is an important consideration in understanding the mechanism of the metal-insulator transition in these pyrochlore ruthenium oxides. It was noted however that the A cation radius alone could not explain the observed nonlinear crossover from metallic to insulating regime \cite{Laurita2019}, but that the stereochemical activity of the \ce{Bi$^{3+}$} 6$s^2$ lone pair electrons leads to local disorder and indirectly increases the \ce{Ru-O-Ru} covalency driving a metallic state in \ce{Bi2Ru2O7}\cite{PhysRevB.72.035124,Avdeev200224,Shoemaker2011,Laurita2019}. 

Metallic \ce{Bi2Ru2O7} is widely used as a thick film resistor\cite{Carcia19825282,Pike19775152,Stanimirovic2020600}, and has further applications as a catalyst\cite{Horowitz19831851}, a cathode material for solid oxide fuel cells \cite{Takeda2002969,Esposito2006}, an electrode for pyrochlore based dielectric material such as \ce{Bi$_{1.5}$Zn$_{1.0}$Nb$_{1.5}$O$_7$}\cite{Nakajima2014} and has been predicted to host fragile band topology \cite{PhysRevX.10.031001}. Although the electrical properties of polycrystalline films have been previously studied in detail\cite{Carcia19825282,Rehak1984647} there are limited studies dedicated to heteroepitaxial \ce{Bi2Ru2O7} thin film growth focusing on (111) oriented heterostructures prepared by chemical vapor deposition methods\cite{Nakajima2014,Chiba2019471}. Pulsed laser deposition (PLD) is a physical vapor deposition method which is well suited to the deposition of high quality oxide thin films and has been used to obtain epitaxial pyrochlore films\cite{Ito2021}, however the high volatility of bismuth and ruthenium makes the stoichiometric transfer of \ce{Bi2Ru2O7} challenging.

In this report we have studied the heteroepitaxial synthesis of pyrochlore \ce{Bi2Ru2O7} thin films by PLD on Y-stabilized \ce{ZrO2} (YSZ) single crystals with two different orientations, (001) and (111). By tuning the deposition conditions, a narrow growth region has been identified where single phase, highly-oriented films could be obtained. The electrical and optical conductivities were measured, and the carrier concentration, obtained from Hall effect measurement, together with the observed optical transition is in good agreement with the calculated electronic structure. The high quality of the thin film samples will therefore allow for further experimental studies to understand the rich electronic structure embedded in this system.

\section{Methods}
\subsection{Experimental methods}
Thin film deposition was carried out using a polycrystalline Bi$_2$Ru$_2$O$_7$ source target synthesized following a previously reported solid state method\cite{Avdeev200224}. Stoichiometric quantities of the dried binary oxides, Bi$_2$O$_3$ and RuO$_2$, were mixed, ground, pressed and calcined at 750 \textcelsius{} for 24 hours in air. The resultant pellets were ground, pressed and fired once again in air at 900 \textcelsius{} for 24 hours, this process was repeated at 950 \textcelsius{} for 72 hours with a final sinter at 975 \textcelsius{} for 24 hours to form a dense ceramic pellet. The phase purity of the obtained target was confirmed by X-ray diffraction (XRD, X'Pert Pro, Panalytical, Co $K_{\alpha 1}$ radiation) and a lattice parameter $a$ = 10.2972(2) \AA{} was obtained in agreement with previous values reported in the literature, 10.2934(1) \AA{}\cite{Kanno1993106} and 10.28974(3) - 10.28976(7) \AA{}\cite{Avdeev200224}. The O Wyckoff position at the 48$f$ site (x, 1/8, 1/8) was found to be x = 0.315(2) which is comparable to the reported figures of 0.32652(6) - 0.32649(6) \cite{Avdeev200224}. Thin films were grown on (001) and (111) oriented 8 mol.\% Y-stabilized zirconia (YSZ) single crystal wafers from Crystal GmbH. The deposition was carried out in a PVD high vacuum chamber with a base pressure of $1\cdot 10^{-7}$ Torr using a KrF excimer laser ($\lambda$ = 248 nm) with a fixed repetition rate of 2 Hz.  The substrate surfaces were prepared by annealing in air at 1150 \textcelsius{} for 3 hours prior to deposition to achieve atomically flat surfaces which were confirmed by atomic force microscopy. A range of substrate temperatures ($T_S$ = 675 -- 750 \textcelsius{}), oxygen partial pressures ($P_{O_2}$ = 1 -- 50 mTorr), and laser energy densities ($J$ = 1.0 -- 1.6 J/cm$^2$) were explored to determine the optimum growth conditions. After deposition the samples were cooled in the same oxygen partial pressure at a rate of 10 \textcelsius{}/minute. The quality (phase and mosaicity) of the films was assessed using a high-resolution four-circle diffractometer (Smartlab, Rigaku, 16 kW rotating anode Cu $K_{\alpha 1}$ radiation). The thickness, $d$, was measured using X-ray reflectometry ($d<$ 170 nm) and profilometry (Ambios XP – 200 stylus profilometer, $d >$ 170 nm). The electrical transport properties were measured in the Van der Pauw configuration \cite{VanderPauw1958} with a 14 T Quantum Design Dynacool ETO using In solder, Au wires and an AC excitation with an amplitude of 0.5 mA and a frequency of 9.15527 Hz. The antisymmetric component of the transverse Hall magnetoresistance was used to calculate the carrier concentration of the films. The optical properties of the \ce{Bi2Ru2O7} films and YSZ substrate were measured using spectroscopic ellipsometry (J. A. Woollam M200UI, wavelength range 240 -- 1700 nm (0.73 -- 5.14 eV)) with photon energies between 0.74 -- 5.86 eV) at three different angles (65, 70 and 75 \textdegree). 

\subsection{Computational methods}
Plane-wave based periodic density functional theory calculations were performed using VASP\cite{VASP} with the PBE functional\cite{PBE}. The projector augmented wave approach\cite{PAW} was used with the inclusion of spin-orbit coupling. Calculations were performed in the primitive unit cell with a 32$\times$32$\times$32 k-point grid and a plane-wave cutoff energy of 550 eV. Prior to the calculation of density of states and band structure, the lattice parameters and atomic positions were optimized until all forces fell below 0.001 eV \AA{}$^{-1}$.

\section{Results and discussion}

The \ce{A2B2O6O\textquotesingle} pyrochlore structure presented in Fig. 1(a) is a face-centered cubic structure which can be viewed as independent interpenetrating \ce{A2O\textquotesingle} and \ce{B2O6} sublattices based on a 3-dimensional framework of corner-sharing \ce{BO6} octahedra that are linked into zigzag chains\cite{Subramanian198355,Kennedy1996261,Kanno1993106,Yamamoto1994372}. The larger A cations occupy the 16$d$ sites at (1/2, 1/2, 1/2) in eight-fold scalenohedral coordination where each A cation is at the centre of an octahedron of \ce{O} squashed along the $<$111$>$ axis and capped by two axial \ce{O\textquotesingle} atoms slightly closer to the A cation. The smaller B cations occupy the 16$c$ sites at (0, 0, 0) assuming trigonal antiprism coordination with the anions, while \ce{O} and \ce{O\textquotesingle} occupy the 48$f$ (x, 1/8, 1/8) and 8$b$ (3/8, 3/8, 3/8) sites respectively in the $Fd\bar{3}m$ space group symmetry. The defect pyrochlore \ce{Bi2Ru2O7} is formed when there are disordered oxygen vacancies on the \ce{O\textquotesingle} 8$b$ site. The B cation sublattice is composed of alternating triangular and kagome structural motifs along the [111] direction, and tight-binding calculations have shown that a two-dimensional kagome lattice with spin-orbit coupling can potentially exhibit topological insulator behavior\cite{PhysRevB.80.113102}.

\subsection{Electronic structure}
The electronic band structure along selected high symmetry directions of the Brillouin zone and density of states (DOS) of \ce{Bi2Ru2O7} are shown in Fig. 1(b) and (c), where the reference energy of the vertical axis is the Fermi energy, $E_{\textnormal{F}}$. The electronic bands positioned between -8 eV and -2 eV (Fig. 1(b)) are composed primarily of \ce{O} 2$p$ states bonding with \ce{Ru} $d$ orbitals at the lower levels and with a \ce{Bi} 6$s$ contribution in the higher energy region with corresponding DOS shown in Fig. 1(c). The Fermi level lies between two localized DOS peaks attributed to antibonding Ru $d$ states which are strongly hybridized with O and Bi $p$ states, consistent with previously reported electronic structure calculations using LDA\cite{Ishii2000526}, PBEsol\cite{Laurita2019} and tight-binding linear muffin-tin orbital method\cite{Avdeev200224}. It is also worth noting some mixing of the Bi $s$ states exists in the $d$ bands above the Fermi level especially around the $\Gamma$ and L points. This leads to an almost semimetallic state with four bands crossing $E_{\textnormal{F}}$, two small electron pockets along the $\Gamma-$K and W$-$L directions and two hole pockets along the $\Gamma-$X and K/U$-$X paths.  The bandwidth of the Ru $t_{2g}$ block bands around the Fermi level at the $\Gamma$ point is ~2.5 eV which is consistent with reported values for LDA \cite{Ishii2000526} and PBEsol\cite{Laurita2019}. 

\begin{figure*}
    \centering
    \includegraphics[width=\textwidth,height=10cm]{"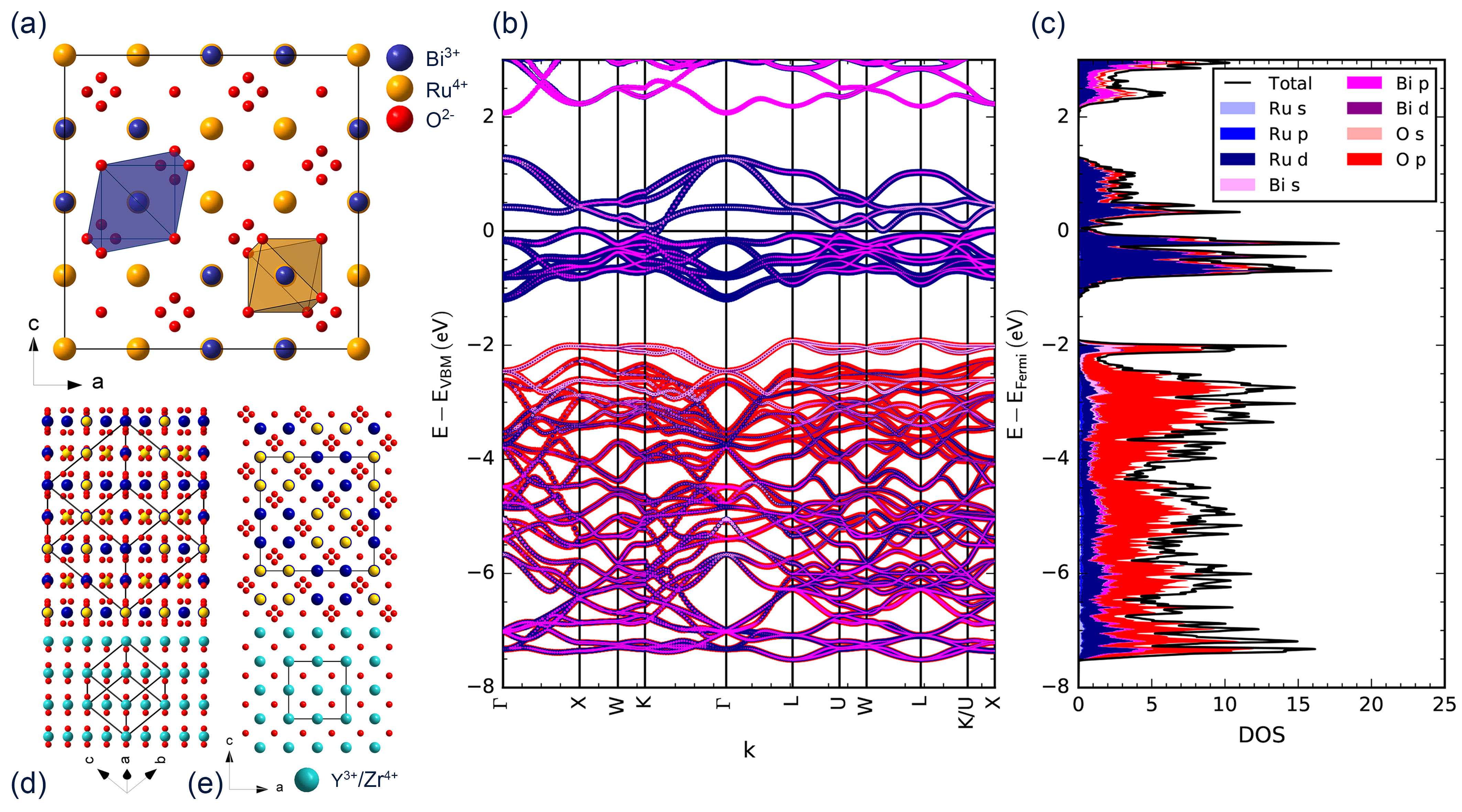"}
    \caption{(a) Pyrochlore Bi$_2$Ru$_2$O$_7$ crystal structure. (b) Bi$_2$Ru$_2$O$_7$/YSZ heterostructure viewed towards the (2$\overline{1} \overline{1}$) plane with the (111) along the vertical axis. (c) Bi$_2$Ru$_2$O$_7$/YSZ heterostructure viewed towards the (010) plane with (001) along the vertical axis. (d) Electronic band structure of Bi$_2$Ru$_2$O$_7$ calculated by density functional theory. (e) Partial densities of states (DOS) of the constituent elements.}
\end{figure*}

\subsection{Thin film growth}
Fluorite YSZ substrates were selected due to the crystal structure similarity and low lattice mismatch of 0.2 \%, to the Bi$_2$Ru$_2$O$_7$ pyrochlore phase as illustrated for (111) and (001) orientations in Fig. 1(d) and (e). Optimal growth was achieved at a temperature of 710 \textcelsius{} under an oxygen partial pressure of 15 mTorr and a laser fluence of 1.6 J/cm$^2$ giving a growth rate of 0.205 \AA/pulse  for the (001) oriented film and 1.4 J/cm$^2$ giving a growth rate of 0.167 \AA/pulse for the (111) oriented film. The relatively high oxygen partial pressure and moderate growth temperature were found necessary both to avoid reduction of the films and to minimize variation in relative stoichiometry of the two volatile cations which led to splitting of the Bragg peaks in the XRD patterns indicative of a stoichiometric inhomogeneity. Higher oxygen pressure was observed to result in an additional Bi$_3$Ru$_3$O$_{11}$ impurity phase. Target polishing was required between depositions to avoid stoichiometric variation in the thin film composition.
\begin{figure}
    \includegraphics[width=\columnwidth/2]{"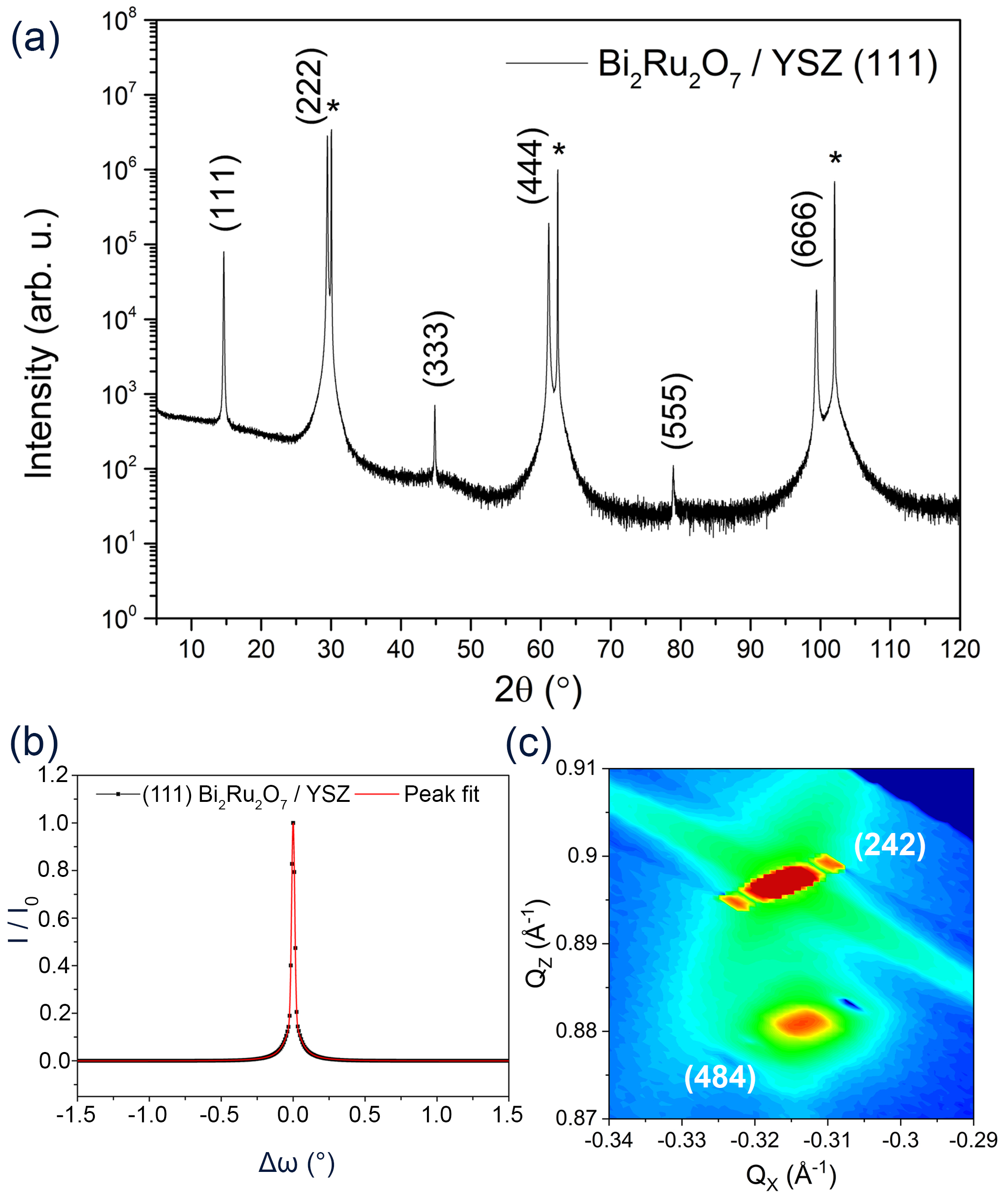"}
    \caption{(a) $\theta$-2$\theta$ X-ray diffraction scan for optimized growth of (111) oriented Bi$_2$Ru$_2$O$_7$/YSZ heterostructure. The film reflections are labelled and the (*) indicates the substrate peaks. (b) Rocking curve of (222) reflection of (111) oriented Bi$_2$Ru$_2$O$_7$/YSZ heterostructure (black line and circle) with pseudo-Voigt peak fit (red line). (c) Reciprocal space map showing the asymmetric Bi$_2$Ru$_2$O$_7$ (484) and the YSZ (242) reflections.}
\end{figure}

The out-of-plane $\theta-2\theta$ XRD pattern for Bi$_2$Ru$_2$O$_7$ on (111) YSZ under optimized growth conditions, presented in Fig. 2(a), shows that the obtained film is single phase and follows the orientation of the substrate with only $hhh$ Bragg reflections observed. The presence of strong odd $hhh$ reflections confirms the cation ordering in the pyrochlore structure rather than the disordered defective fluorite which was obtained in zirconate pyrochlore heterostructures grown by PLD\cite{O'Sullivan2016347}. An out-of-plane lattice parameter of $a$ = 10.47936 \AA{} for the (111) oriented film was observed to be larger than the reported bulk value\cite{Avdeev200224} and likely due to slightly off-stoichiometric cation transfer due to the high vapor pressure of Bi and Ru, and variation in anion content. The quality of the heterostructure was assessed by a rocking curve measurement on the (111) Bi$_2$Ru$_2$O$_7$ Bragg reflection shown in Fig. 2(b). The peak exhibits a broad base with a narrow component which can be fitted using a pseudo-Voigt profile with Gaussian and Lorentzian widths of $w_G$ = 0.025(3)\textdegree{} and $w_L$ = 0.121(2)\textdegree{} respectively. The high intensity, narrow component is caused by the diffraction of (111) planes and the broad, lower intensity component is attributed to diffuse scattering originating from interface roughness, high dislocation density and strain in quasi-epitaxial thin films\cite{Blasing2009}. The crystalline quality of these films is superior to the previously reported (111) oriented \ce{Bi2Ru2O7} films which presented rocking curve peak widths of 0.2\textdegree{}\cite{Nakajima2014,Chiba2019471}. The reciprocal space map collected around the (242) YSZ Bragg reflection is presented in Fig. 2(c) and shows that the (111) oriented \ce{Bi2Ru2O7} film is strained to the substrate. An in-plane lattice parameter of $a$ = 10.420 \AA{} was observed which is close to the out-of-plane component and the grown film can be considered a close equivalent to a bulk \ce{Bi2Ru2O7} single crystal.
\begin{figure}
    \includegraphics[width=\columnwidth/2]{"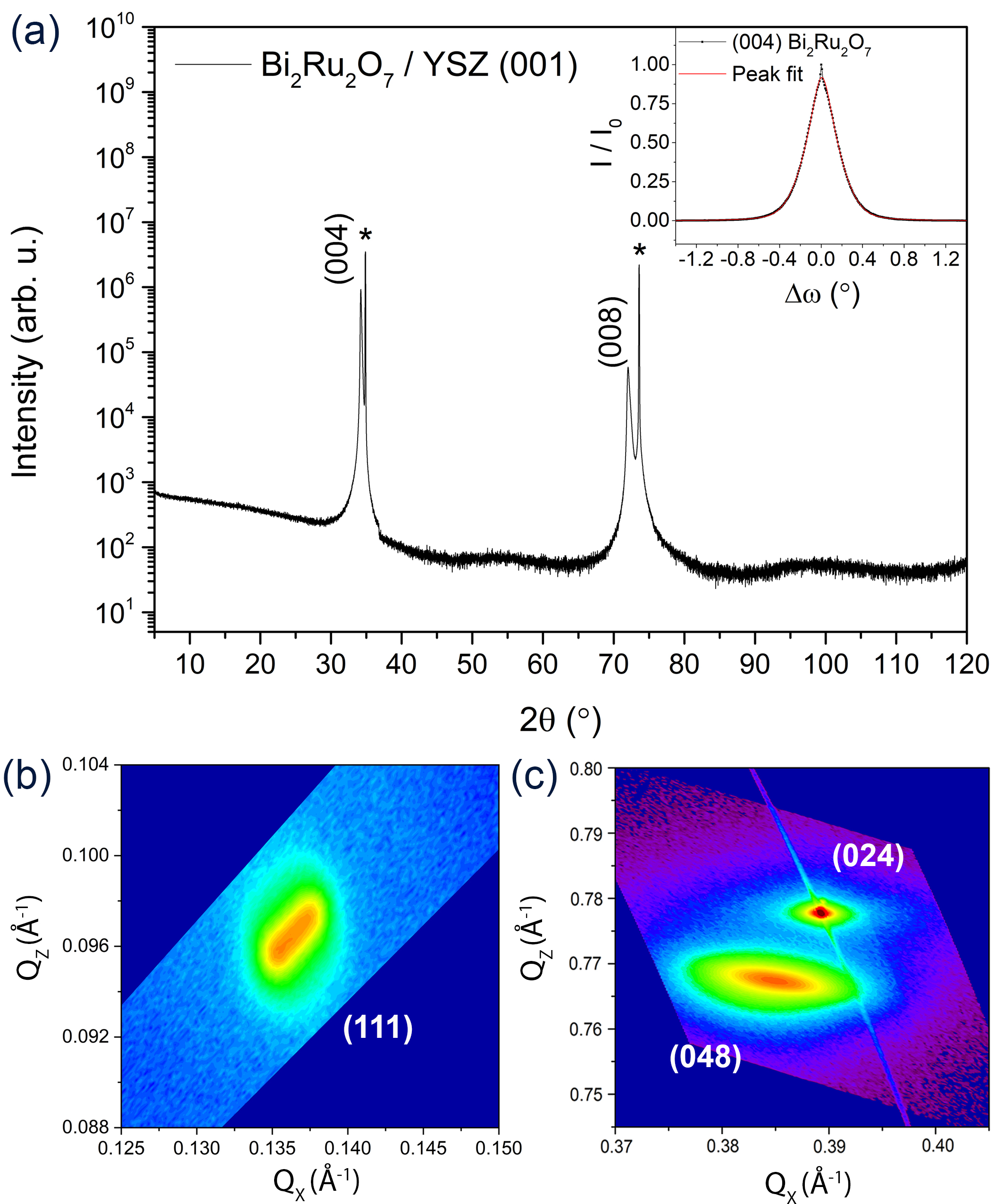"}
    \caption{(a) $\theta$-2$\theta$ X-ray diffraction scan for optimized growth of (001) oriented Bi$_2$Ru$_2$O$_7$/YSZ heterostructure. The (00$l$) film reflections are labelled and the (*) indicates the substrate peaks. Inset shows the rocking curve of the (004) reflection of (001) oriented Bi$_2$Ru$_2$O$_7$/YSZ heterostructure (black line and circle) with pseudo-Voigt peak fit (red line). (b) Reciprocal space map showing the asymmetric Bi$_2$Ru$_2$O$_7$ (111) reflection of the (001) oriented Bi$_2$Ru$_2$O$_7$/YSZ heterostructure confirming the cation ordering in the pyrochlore crystal structure. (c) Reciprocal space map showing the asymmetric Bi$_2$Ru$_2$O$_7$ (048) and the YSZ (024) reflections in the (001) oriented Bi$_2$Ru$_2$O$_7$ film.}
\end{figure}
The sample deposited on the (001) YSZ substrate under the optimized temperature and partial pressure yielded a phase pure \ce{Bi2Ru2O7} film, as shown in the out-of-plane XRD pattern in Fig. 3(a). The observation of ($00l$) \ce{Bi2Ru2O7} Bragg reflections alone confirm that it is highly oriented and the extracted lattice parameter, $a$ = 10.47388 \AA{}, which is close to that of the (111) oriented film, indicates a comparable stoichiometric transfer for both orientations. The rocking curve measured on the (004) \ce{Bi2Ru2O7} reflection is shown in the inset of Fig. 3(a) and a fit to a pseudo-Voigt function resulted in widths of $w_G$ = 0.377(2)\textdegree{} and $w_L$ = 0.323(5)\textdegree. In this case the two components are almost equal and the rocking curve is broader than that obtained for the (111) oriented film. The inferior crystallinity of the (001) oriented sample could be due to the mixed cation composition in the (111) planes of the structure (Fig. 1(e)) compared to uniquely cationic layers alternating with complete layers of anions in the (001) planes (Fig. 1(d)) which may introduce stacking faults. It could also originate from the crystallization of the defective fluorite structure where the Bi and Ru atoms occupy the same crystallographic site. Since the ($002n$) reflections are forbidden by symmetry, it is not possible to distinguish (001) oriented \ce{Bi2Ru2O7} pyrochlore and defective fluorite out-of-plane. The cation ordering in the film was confirmed with the observation of the (111) \ce{Bi2Ru2O7} reflection presented in Fig. 3(b) and the film was found to be fully relaxed with an in-plane lattice parameter of $a$ = 10.395 \AA{} extracted from the \ce{Bi2Ru2O7} peak in the reciprocal space map measured around the (024) reflection of the YSZ substrate. 

\subsection{Electrical properties}
Both film orientations show a similar metallic temperature dependence in the electrical resistivity with a small negative temperature coefficient across the measured temperature range (Fig. 4). This cannot be explained by a thermally activated semiconducting mechanism and is similar to previously reported thin film \cite{Nakajima2014,Chiba2019471} and polycrystalline samples prepared in air \cite{Carbonio1999361}. Resistivity values of $\rho$ = 0.54(3) m$\Omega$cm and $\rho$ = 0.43(3) m$\Omega$cm were obtained for the (001) and the (111) oriented layers respectively at 300 K, in close agreement with the values reported in polycrystalline samples \cite{Kanno1993106,Carbonio1999361}, single crystals \cite{PhysRevB.73.193107} and thin films \cite{Nakajima2014,Chiba2019471}. The reduced resistivity observed in the (111) oriented film could be attributed to lower scattering from dislocations as the crystallinity of this film is superior. 

\begin{figure}
    \includegraphics[width=\columnwidth/2]{"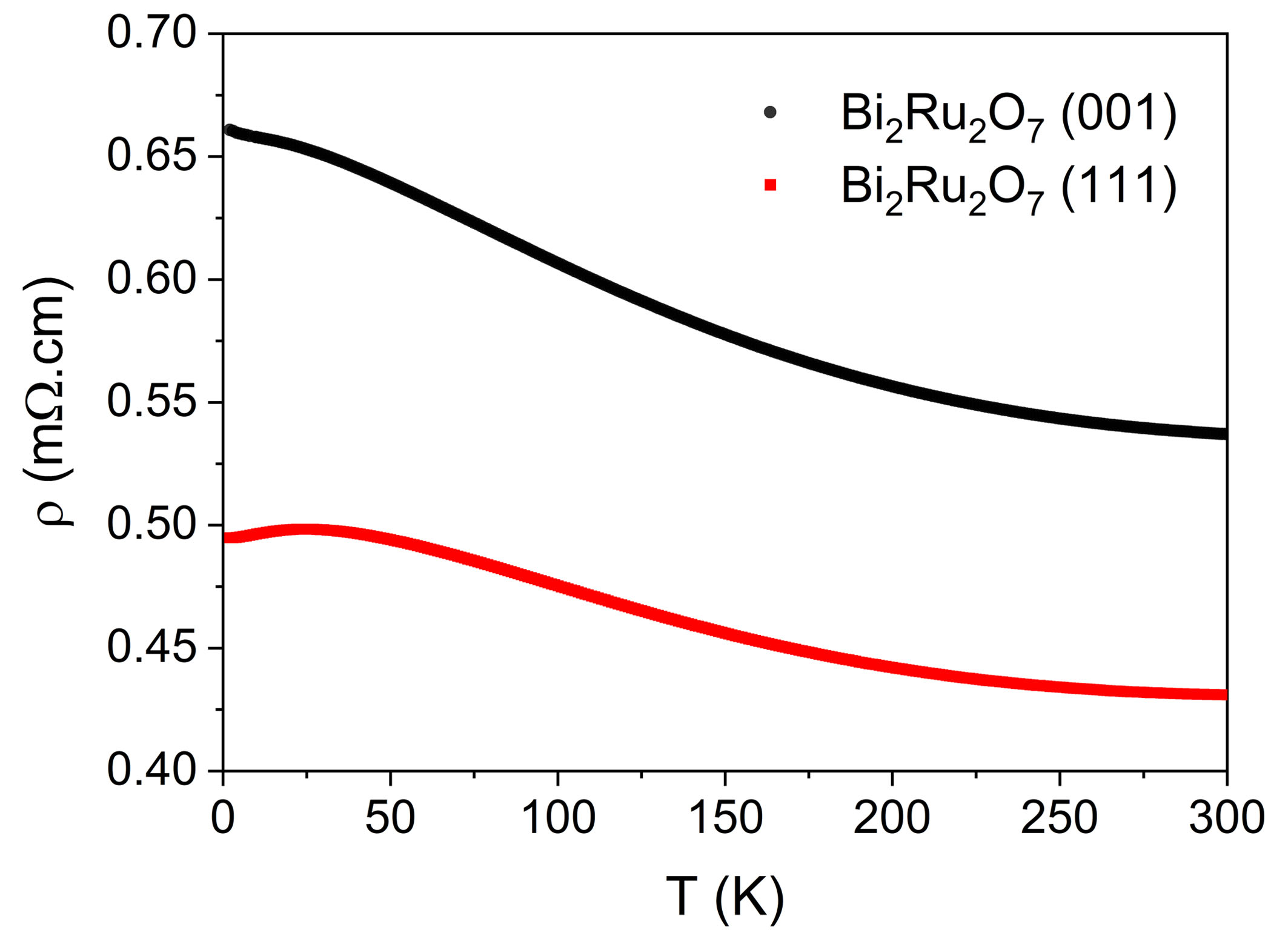"}
    \caption{(a) Temperature dependence of electrical resistivity of (111) oriented (red squares) and (001) oriented (black circles) Bi$_2$Ru$_2$O$_7$/YSZ heterostructures.}
\end{figure}

To our knowledge, the carrier concentration in this compound has only been estimated through spectroscopic measurements \cite{Cox19836221} and the thin film configuration in this study permits measurement of the effective carrier concentration using conventional Hall effect. The magnetic field dependence of the Hall resistance, shown in Fig. 5(a) is linear with a negative slope indicating the majority charge carriers are electrons. The temperature dependence of the effective carrier concentration calculated from the Hall resistance for both (001) and (111) orientations is presented in Fig. 5(b). The samples exhibit an increase in carrier concentration with increasing temperature, differing from the nearly constant carrier concentration expected for a metal or degenerate semiconductor. The carrier concentrations measured at 5 K are 7.5(5)$\cdot 10^{22}$ cm$^{-3}$ and 3.6(2)$\cdot 10^{22}$ cm$^{-3}$ for the (111) and (001) oriented films respectively. These values are close to the carrier concentration extracted from spectroscopic measurements \cite{Cox19836221} and the theoretical carrier concentration of 5.9$\cdot 10^{22}$ cm$^{-3}$ expected for four free electrons per Ru. Carrier concentrations of 2.9(2)$\cdot 10^{23}$ cm$^{-3}$ and 1.20(8)$\cdot 10^{24}$ cm$^{-3}$ were observed at 300 K for the (001) and (111) oriented heterostructures respectively which may arise due to thermal broadening of the Fermi distribution in the higher temperature region. The temperature dependence of the extracted Hall mobility is presented in Fig. 5(c) for both orientations showing an increase in the mobility as the temperature is reduced towards a saturation values of 0.17 and 0.28 cm$^2$V$^{-1}$s$^{-1}$ for the (111) and (001) oriented samples respectively at low temperature. This suggests that the dominant mechanism limiting the mobility is phonon scattering at high temperature and neutral impurities at low temperature. The low mobility values are consistent with the reported conductivity and large effective mass of 2.9 $m_0$ observed in polycrystalline samples \cite{Kanno1993106,Cox19836221}. 

\begin{figure*}
    \centering
    \includegraphics[width=\textwidth,height=5.1cm]{"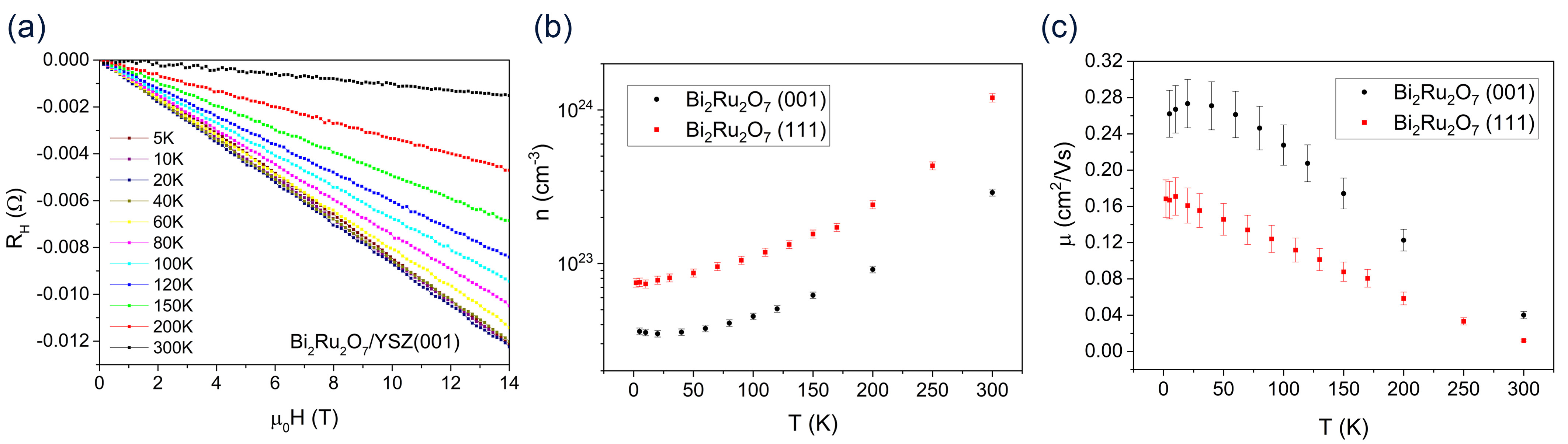"}
    \caption{(a) Hall resistivity of a (001) oriented Bi$_2$Ru$_2$O$_7$/YSZ heterostructure measured in the Van der Pauw geometry. (b) Carrier mobility and (c) carrier concentration of (111) oriented (red squares) and (001) oriented (black circles) Bi$_2$Ru$_2$O$_7$/YSZ heterostructures extracted from Hall effect measurements.}
\end{figure*}

\subsection{Optical properties}
The real $\epsilon_1$ and imaginary $\epsilon_2$ parts of the dielectric function obtained from ellipsometry data for both (001) and (111) orientations are shown in Fig. 6(a) and (b) respectively. The raw ellipsometry data were fitted with a combination of Drude and Lorentz oscillators in a combined model for the substrate and the film to extract the complex dielectric function. A blank substrate was measured to determine the dielectric function and a model using three Tauc-Lorentz oscillators, similar to what is reported for c-\ce{ZrO2} \cite{Synowicki2004248}, was fixed together with the film thickness in the heterostructure models. The film layer models were constructed from a Drude oscillator combined with a series of five Lorentz oscillators. The real components of the optical conductivity spectra $\sigma_1$ were calculated and shown in Fig. 6(c) together with the optical conductivity extracted from the DFT calculation and that of a previous study performed on a polycrystalline sample \cite{PhysRevB.72.035124}. The reduced plasma energy $E(\omega_p$), defined as the energy at which the real part of the complex dielectric function $\epsilon_1$ is equal to zero, was measured to be 1.17 eV for the (111) oriented film and 1.34 eV for the (001) oriented sample. These values are comparable to the reported electron energy-loss spectral (EELS) peak of 1.25 eV for a polycrystalline sample which was associated with excitation of the conduction-electron surface plasmon \cite{Cox19836221}. The red shift in the plasma energy from the (001) to the (111) oriented film together with the elevated carrier concentration observed in the magnetotransport measurements in the latter film and must indicate a coincident increase in electron effective mass which is inversely proportional to the plasma frequency. Despite the superior crystalline quality of the (111) heterostructure, demonstrated by the narrow rocking curve, the Fermi surface of the (111) planes could have a higher concentration of carriers together with a diminished carrier mobility compared with the (001) if either the effective mass was sufficiently large or the scattering time was sufficiently enhanced in an electronically anisotropic crystal structure.
\begin{figure*}
    \centering
    \includegraphics[width=\textwidth,height=5.0cm]{"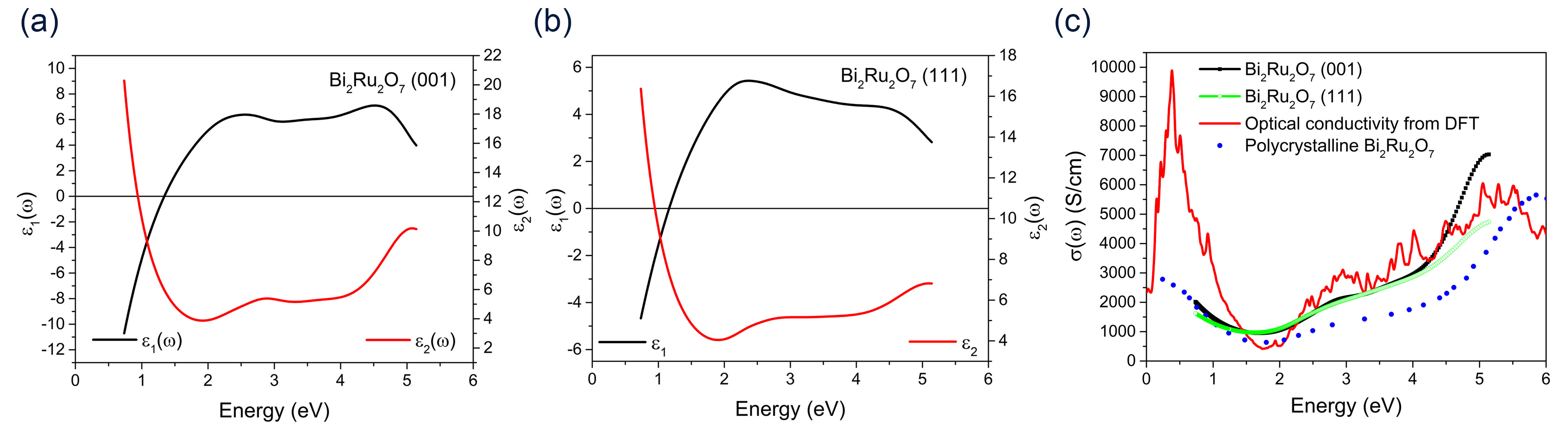"}
    \caption{Ellipsometry measurements showing the real $\epsilon_1$ and imaginary $\epsilon_2$ parts of the dielectric function for (a) (001) and (b) (111) oriented Bi$_2$Ru$_2$O$_7$/YSZ heterostructures. (c) Calculated optical conductivity $\sigma(\omega)$ for (001) and (111) oriented Bi$_2$Ru$_2$O$_7$/YSZ heterostructures compared with the optical conductivity extracted from the DFT calculations and that of a polycrystalline sample studied by Lee $et$ $al.$ \cite{PhysRevB.72.035124}.}
\end{figure*}

Electronic band structure calculations indicate deep bands composed of O 2$p$ bonding with Ru $d$ orbitals and some contribution from Bi 6$s$ at levels well below $E_{\textnormal{F}}$. Closer to $E_{\textnormal{F}}$ the bands consist mainly of a Ru 4$d$ $t_{2g}$ block with a bandwidth of 2.5 eV at $\Gamma$. The $E_{\textnormal{F}}$ lies between two localized density of states peaks indicating that the material is semimetallic, this corresponds to Ru 4$d$ states which are hybridized with Bi 6$p$ states via the anion network in agreement with more recent theoretical studies \cite{Hsu1988792,Ishii2000526}. Epitaxial films of (001) and (111) oriented Bi$_2$Ru$_2$O$_7$/YSZ heterostructures have been grown by PLD. Both films crystallized with the ordered pyrochlore structure with lattice parameters that were larger than the reported bulk figures \cite{Avdeev200224}. The (111) oriented film had a superior crystallinity and was strained to the substrate. The films were weakly metallic with resistivity profiles in agreement with previously reported electrical measurements \cite{Carbonio1999361}. They had low electron mobilities, consistent with optical spectroscopy reports of high carrier effective masses \cite{Cox19836221,Kanno1993106}, and they had carrier concentrations in agreement with the theoretical concentration with thermally activated carriers at higher temperature. The reduced plasma energies extracted from the real component of the dielectric function were close to the EELS surface plasmon excitation energy previously reported \cite{Cox19836221}. The spectral features of the experimental optical conductivity of the films were reproduced in the DFT computed model. Differences in the magnetotransport parameters and the optical properties between the two orientations which cannot be explained by differences in crystallinity indicate a degree of electrical and optical anisotropy in the electronic structure. 

\section{Conclusions}
In summary, we have grown epitaxial heterostructures of pyrochlore Bi$_2$Ru$_2$O$_7$ along [111] and [001] directions of YSZ single crystals by PLD. The structural properties, crystalline quality and strain in the heterostructures have been examined by XRD. The electrical resistivity has been measured and the carrier type, mobility and concentration have been determined by Hall effect measurement as a function of temperature. The optical conductivity and reduced plasma energy have been determined by ellipsometry measurements. The transport and optical parameters have been compared for the two orientations and a degree of anisotropy in the crystal structure is evident. The measured optical conductivities have also been compared with the DFT calculated optical conductivity and the observed optical transitions have been reproduced at coincident energies in the computed electronic structure. Investigation of the effect of Bi off-centering on the electronic structure and optical conductivity would give useful insights into what drives the metallicity in this system. The ability to access distinct structural motifs such as the kagome lattice which is peculiar to the B cation planes of the (111) orientation of the pyrochlore crystal structure could lead to new and interesting physical properties relating to the symmetry of this structure. Recent computational studies into the 3-dimensional pyrochlore niobate systems suggest that topological phases can be tuned by shifting the A cation away from the centrosymmetric position along the [111] axis\cite{Zhang2019, PhysRevB.99.201105}, indicating the potential of this crystal system to demonstrate exciting symmetry-protected band structure features.
\nocite{*}

\section*{Acknowledgements}
This paper is dedicated to Sir A.K. Cheetham in celebration of his 75th birthday.

The authors would like to acknowledge funding from the Engineering and Physical Sciences Research Council (EPSRC) Programme Grants (EP/N004884 and EP/V026887).  The X-ray diffraction facility used to characterize the films was supported by EPSRC (grant EP/P001513/1). We thank the Leverhulme Trust for funding via the Leverhulme Research Centre for Functional Materials Design. MOS acknowledges financial support from the EPSRC as part of a Daphne Jackson Fellowship.

\section*{Authors declaration}
\subsection*{Conflict of interest}
The authors have no conflicts to disclose.
\subsection*{Author contributions}
Marita O'Sullivan: Conceptualization; Formal analysis; Investigation; Writing - Original Draft; Visualization. Matthew S. Dyer: Formal analysis; Investigation; Writing - Review \& Editing; Visualization. Michael W. Gaultois: Writing - Review \& Editing. John B. Claridge: Conceptualization; Writing - Review \& Editing. Matthew J. Rosseinsky: Conceptualization; Resources; Writing - Review \& Editing; Project administration; Funding acquisition. Jonathan Alaria: Formal analysis; Writing - Original Draft.
\section*{Data availability}
The data that support the findings of this study are openly available in 
\section*{References}
\bibliography{BRO}

\end{document}